\documentstyle [12pt,twoside]{article}

\newcommand{\bq}{\begin{equation}}
\newcommand{\ba}{\begin{eqnarray}}
\newcommand{\eq}{\end{equation}}
\newcommand{\ea}{\end{eqnarray}}

\def\c{\raisebox{.4ex}{$\chi$}}
\def\d{\delta}

\def\s{\sigma}

\def\D{\Delta}

\def\G{\Gamma}

\def\L{\Lambda}

\def\bo{{\raise.15ex\hbox{\large$\Box$}}}
\def\bob{{\lower.2ex\hbox{\large$\Box$}}}

\def\TH{{\raise.2ex\hbox{$\displaystyle \bigodot$}\mskip-4.7mu \llap H \;}}
\def\face{{\raise.2ex\hbox{$\displaystyle \bigodot$}\mskip-2.2mu \llap {$\ddot
        \smile$}}}

\def\Hat#1{\rlap{\kern.10em$\widehat{\phantom G}$}#1}
\def\HAt#1{\rlap{\kern.05em$\widehat{\phantom G}$}#1}

\def\cap#1{\rlap{\kern.1em$\widehat{\phantom{G\vrule height.8em}}$}#1{}}
\def\Cap#1{\rlap{\kern.05em$\widehat{\phantom{G\vrule height.8em}}$}#1{}}

\def\leftrightarrowfill{$\mathsurround=0pt \mathord\leftarrow \mkern-6mu
        \cleaders\hbox{$\mkern-2mu \mathord- \mkern-2mu$}\hfill
        \mkern-6mu \mathord\rightarrow$}
\def\overleftrightarrow#1{\vbox{\ialign{##\crcr
        \leftrightarrowfill\crcr\noalign{\kern-1pt\nointerlineskip}
        $\hfil\displaystyle{#1}\hfil$\crcr}}}

\def\frac#1#2{{\textstyle{#1\over\vphantom2\smash{\raise.20ex
        \hbox{$\scriptstyle{#2}$}}}}}

\def\sfrac#1#2{{\vphantom1\smash{\lower.5ex\hbox{\small$#1$}}\over
        \vphantom1\smash{\raise.4ex\hbox{\small$#2$}}}}
\def\bfrac#1#2{{\vphantom1\smash{\lower.5ex\hbox{$#1$}}\over
        \vphantom1\smash{\raise.3ex\hbox{$#2$}}}}
\def\afrac#1#2{{\vphantom1\smash{\lower.5ex\hbox{$#1$}}\over#2}}

\catcode`@=11
\def\underline#1{\relax\ifmmode\@@underline#1\else
        $\@@underline{\hbox{#1}}$\relax\fi}
\catcode`@=12

\def\nis{\nointerlineskip}
\def\Abar{\vbox{\nis\moveright.33em\vbox{
        \hrule width.35em height.04em}\nis\kern.05em\hbox{$A$}}{}}
\def\Dbar{\vbox{\nis\moveright.20em\vbox{
        \hrule width.50em height.04em}\nis\kern.05em\hbox{$D$}}{}}
\def\Gbar{\vbox{\nis\moveright.20em\vbox{
        \hrule width.50em height.04em}\nis\kern.05em\hbox{$G$}}{}}
\def\mbar{\vbox{\nis\moveright.15em\vbox{
        \hrule width.60em height.04em}\nis\kern.05em\hbox{$m$}}{}}
\def\Rbar{\vbox{\nis\moveright.20em\vbox{
        \hrule width.50em height.04em}\nis\kern.05em\hbox{$R$}}{}}
\def\Vbar{\vbox{\nis\moveright.05em\vbox{
        \hrule width.60em height.04em}\nis\kern.05em\hbox{$V$}}{}}
\def\Xbar{\vbox{\nis\moveright.20em\vbox{
        \hrule width.60em height.04em}\nis\kern.05em\hbox{$X$}}{}}
\def\thetabar{\vbox{\nis\moveright.15em\vbox{
        \hrule width.30em height.04em}\nis\kern.05em\hbox{$\theta$}}{}}
\def\Lambdabar{\vbox{\nis\moveright.25em\vbox{
        \hrule width.35em height.04em}\nis\kern.05em\hbox{${\mit\Lambda}$}}{}}
\def\Sigmabar{\vbox{\nis\moveright.25em\vbox{
        \hrule width.50em height.04em}\nis\kern.05em\hbox{${\mit\Sigma}$}}{}}
\def\phibar{\vbox{\nis\moveright.18em\vbox{
        \hrule width.40em height.04em}\nis\kern.05em\hbox{$\phi$}}{}}
\def\chibar{\vbox{\nis\moveright.12em\vbox{
        \hrule width.40em height.04em}\nis\kern.05em\hbox{$\chi$}}{}}
\def\psibar{\vbox{\nis\moveright.23em\vbox{
        \hrule width.40em height.04em}\nis\kern.05em\hbox{$\psi$}}{}}
\def\debar{\vbox{\nis\moveright.18em\vbox{
        \hrule width.35em height.04em}\nis\kern.05em\hbox{$\partial$}}{}}
\def\delbar{\vbox{\nis\moveright.10em\vbox{
        \hrule width.63em height.04em}\nis\kern.05em\hbox{$\nabla$}}{}}

\newskip\humongous \humongous=0pt plus 1000pt minus 1000pt

\newif\ifdtup

\def\begintitle#1#2#3#4
        {\begin{titlepage}
         \centerline{#1 \hfill MdDP-PP-88-#2}
         \begin{center}\vglue .7in
         {\large\bf #3}\\].7in(
         {\bf #4}\\
         {\it Department of Physics and Astronomy}\\
         {\it University of Maryland, College Park, MD 20742}\\].7in(
         {\bf ABSTRACT}\\
         \end{center}
         \begin{quotation}}
\def\endtitle
         {\end{quotation}
          \end{titlepage}
          \newpage}

\begin{document}
\begin{titlepage}
\begin{large}\par\noindent
\end{large}

\vspace{12pt}
\begin{center}
\begin{Large}
\renewcommand{\thefootnote}{\fnsymbol{footnote}}
\bf{
Short Times Characterisations of Stochasticity
in Nonintegrable Galactic Potentials
}\footnote[1]
{To appear in: {\em Astronomy and Astrophysics}}
\end{Large}

\vspace{12pt}

\begin{large}
\renewcommand{\thefootnote}{\fnsymbol{footnote}}
Henry E. Kandrup\footnote[2]
{Also Department of Physics, University of Florida:
kandrup@astro.ufl.edu}
 and M. Elaine Mahon\\

Department of Astronomy and Institute for Fundamental Theory\\
University of Florida, Gainesville, FL 32611 \\
\end{large}
\vspace{12pt}
\end{center}
This paper proposes a new, potentially useful, way in which to characterise
the degree of stochasticity exhibited by orbits in a fixed time-independent
galactic potential. This approach differs from earlier work involving the
computation of Liapounov exponents in two ways, namely (1) by focusing on the
statistical properties of ensembles of trajectories, rather than individual
orbits, and (2) by restricting attention to the properties of these ensembles
over time scales shorter than the age of the Universe, $t_{H}$. For many
potentials, generic ensembles of initial conditions corresponding to stochastic
orbits will evolve relatively quickly towards a time-independent invariant
measure ${\G}$, which is arguably the natural unit to
consider if one is interested in self-consistent equilibria. The basic idea
proposed here is to compute short time Liapounov characteristic numbers
${\chi}({\D}t)$ over time intervals ${\D}t$ for orbits
in an ensemble that samples this invariant measure, and to analyse the overall
distribution of these ${\chi}$'s. This is done in detail for one model
potential, namely the sixth order truncation of the Toda lattice potential.
One especially significant conclusion is that time averages and
ensemble averages coincide, so that the form of the distribution of short time
${\chi}({\D}t)$'s for such an ensemble is actually encoded in the calculation
of ${\chi}(t)$ for a single orbit over long times $t{\;}{\gg}{\;}t_{H}$. The
distribution of short time ${\chi}$'s is analysed as a function of the
energy $E$ of orbits in the ensemble and the length of the short time sampling
interval ${\D}t$. For relatively high energies, the distribution is essentially
Gaussian, the dispersion decreasing with time as $t^{-p}$, with an exponent
$0<p<1/2$ that depends on the energy $E$.
\end{titlepage}
\par\noindent
{\bf 1. MOTIVATION}
\vskip .1in
\par \noindent
Dating back to the pioneering work of H\'enon and Heiles (1964) thirty years
ago, considerable attention has focused on the study of orbits in nonintegral
potentials which are thought to reflect the bulk mass distribution of a galaxy
(see, e.g., Martinet \& Pfenniger 1987 or Contopoulos \& Grosbol 1989,
and references cited therein). The Hamiltonian systems associated with these
potentials differ from integrable Hamiltonian systems in that they may admit
both regular and stochastic orbits.
\par
In order to characterise the stochastic orbits, it is useful to introduce the
concept of a Liapounov exponent. One way in which to define such Liapounov
exponents (cf. Pesin 1977) is in a fashion related directly to
the Kolmogorov entropy (cf. Chirikov 1979). However, it quite natural
physically and more convenient computationally to define these exponents
instead in terms of the stability of individual stochastic trajectories
(Bennetin et al 1976) with respect to small perturbations. Specifically,
Liapounov exponents may be defined by the prescription
$${\c}{\;}{\equiv}{\;}\lim _{t\to\infty}{\;}\lim_{{\d}r(0)\to 0}
{\;}{1\over t}{\;}{\rm log}{\Biggl(}
{{\d}r(t)\over {\d}r(0)}{\Biggr)},  \eqno(1)$$
where ${\d}r(0)$ and ${\d}r(t)$ denote respectively the configuration space
deviations of two nearby orbits at times $0$ and $t$. Of particular interest
is the maximal Liapounov exponent corresponding to the most unstable
perturbation, which can be computed numerically by selecting
${\d}r(0)$ at random. In everything that follows, the words ``Liapounov
exponent'' will refer to this maximal Liapounov exponent.
\par
This prescription involves a formal $t\to\infty$ limit. Clearly, however,
one cannot integrate forever, so that, as a practical matter, one is
restricted to computing finite times estimates of $\c$. The crucial question
then is: For how long must one integrate to obtain reasonable estimates?
\par
The precise answer to this question depends on the particular form of the
potential. However, one knows from experience (cf. Contopoulos and
Barbanis 1989) that, as a general rule,
one must typically integrate over extremely long times, say $t{\;}{\ge}
{\;}10^4 t_D$, where $t_D$ denotes a characteristic dynamical, or crossing,
time. Unfortunately, though, $t_D{\;}{\sim}{\;}10^8$ yr for a galaxy like the
Milky Way, so that this entails an integration for a period of time that is
orders of magnitude longer than the age of the Universe. For this reason,
the Liapounov exponent $\c$ does not provide a useful characterisation of
the instability of individual trajectories on short time scales
${\le}{\;}t_{H}$.
\par
Motivated by this realisation, a number of different workers, notably Udry
and Pfenniger (1988), have sought instead to define shorter time analogues
of the Liapounov exponent which could provide a characterisation of the
stochasticity of individual orbits on time scales ${\le}{\;}t_H$. This is an
extremely interesting idea. However, these authors have not yet fully
explored the potential implications of such an approach.
\par
Most of the work hitherto on the problem of stochasticity in galactic
potentials has been formulated in terms of the properties of individual
orbits. However, even though it is convenient computationally to study such
individual trajectories, there are solid physical motivations for working
instead with ensembles of orbits. Perhaps the most obvious point is that
individual orbits of stars are not accessible observationally: all that
one can detect is an instantaneous snapshot of the overall surface brightness
of a galaxy which, one hopes, traces the distribution of mass. Another
obvious point is that one is really compelled to consider ensembles of
orbits, rather than individual trajectories, if one is interested in
incorporating stochastic orbits into a self-consistent model using
Schwarzschild's (1979) method, or any variant thereof.
\par
The object of this paper is to propose an alternative approach to the
analysis of stochastic orbits which confronts these basic difficulties.
Specifically, the aim is to provide useful characterisations of stochasticity
appropriate for ensembles of orbits, focusing exclusively on relatively
short time scales ${\le}{\;}t_H$.
\par
A consideration of stochasticity for individual orbits on short time scales
is a very complicated proposition and, for this reason, has been
avoided by most nonlinear dynamicists. However, even though the transient
behavior of individual orbits is quite complex, the statistical
properties of ensembles of orbits can, nevertheless, exhibit striking
regularities which facilitate a relatively simple characterisation.
\par
Three basic facts, described in detail below, facilitate this general
approach. The first is that, for many potentials, including the model system
analysed in this paper (Kandrup \& Mahon 1994), generic ensembles
of initial conditions corresponding to stochastic orbits will evolve on a
relatively short time scale towards a particular distribution which is
time-independent or, if not strictly time-independent, only exhibits
subsequent variability on very long time scales. The approach towards a
strictly time-independent invariant measure is in fact guaranteed for
Hamiltonian systems in which the stochastic orbits are ergodic, although there
is no reason theoretically to expect that the time scale associated with this
approach is short.
\par
The second fact is that, in many contexts, it is this
invariant, or near-invariant, measure which
constitutes the natural object in terms of which to analyse the statistical
properties of stochasticity. In particular, it is reasonable to consider an
ensemble of stochastic orbits that samples the invariant measure, and then
compute the Liapounov characteristic numbers for the orbits in that
ensemble over time intervals ${\D}t{\;}{\le}{\;}t_{H}$. The statistical
properties of the resulting {\it distribution of Liapounov characteristic
numbers} then provides a useful characterisation of the overall degree of
stochasticity for the ensemble.
\par
The final fact is that, for an ensemble that samples the invariant measure,
ensemble averages and time averages may be equivalent, so that all the
statistical information about stochasticity appropriate for such ensembles
can actually be contained within the standard calculation of a Liapounov
characteristic number $\c(t)$ for a single orbit over very long times $t$.
For the case of a true invariant measure, this equivalence follows from the
ergodic theorem.
\par
These three facts indicate that it is especially natural to compute short
time Liapounov characteristic numbers for the particular case of ensembles of
orbits that sample the invariant measure. However, one can equally well
choose to compute these characteristic numbers for other ensembles. Thus,
for example, one might focus upon an ensemble initially localised in some
small phase space region which
eventually evolves towards the invariant measure, and then compute the
Liapounov characteristic numbers for the orbits in that ensemble. Such a
computation is potentially interesting in that it permits a characterisation
of the overall stochasticity of a ``nonequilibrium'' ensemble as it ``evolves
towards equilibrium,'' which can facilitate useful insights into the
connection between stochasticity and relaxation.
\par
Section 2 of this paper outlines the general tact to be used in providing
short time characterisations of stochasticity for ensembles of orbits.
The remaining sections then focus on an implementation of this approach
for one particular model, namely the sixth order truncation of the
Toda (1967a,b) lattice potential. This specific form was not selected because
of the expectation that it provides an especially good approximation to the
gravitational potential of any particular class of galaxies. Rather, in
the spirit of the pioneering work of H\'enon and Heiles (1964), the aim has
been to choose a potential which, in the sense described in Section 3, appears
to be generic; and to use that potential as a testing ground for various ideas.
\par
Section 3 of the paper first demonstrates the concrete sense in which the
information about short time measures of stochasticity is actually encoded
in the long time calculations of ${\c}(t)$, and then analyses the form of the
distribution of short time Liapounov characteristic numbers associated with
an ensemble of orbits that samples the invariant measure. Section 4 next
outlines one concrete prescription for generating ensembles that sample this
invariant measure by evolving initially localised ensembles of initial
conditions, and then analyses the distribution of Liapounov characteristic
numbers associated with these initially localised ensembles.
Section 5 concludes by examining the form of these distributions
as a function of the sampling interval ${\D}t$.
\vfill
\eject
\vskip .2in
\par\noindent
{\bf 2. TRANSIENT DYNAMICS OF AN ENSEMBLE OF STOCHASTIC}
\par
{\bf ORBITS}
\vskip .1in
\par \noindent
The standard formulation of nonlinear dynamics, as applied to galactic
dynamics, entails a consideration of {\it asymptotic orbital dynamics},
focusing explicitly on the asymptotic, long time behavior of individual
trajectories. The object here is to propose a new approach, {\it transient
ensemble dynamics.} In this alternative approach, the principal focus is
on the statistical properties of ensembles of orbits, rather than the details
of individual trajectories. Moreover, the analysis is restricted entirely
to properties of these orbits on short time scales $\le{\;}t_H$. In this
approach, asymptotic calculations over extremely long time intervals are
eschewed as unphysical, except to the extent that they can provide information
about the behavior of ensembles of orbits on short, astrophysically relevant,
time scales.
\par
As noted already, the transient behavior of individual orbits can be
extremely difficult to characterise, so that a description of stochasticity
in terms of transient orbital dynamics would typically prove quite
complicated. However, there are intuitive reasons to believe that,
nevertheless, the physically relevant aspects of the short time behavior can
be characterised relatively easily if one chooses instead to focus on the
statistical properties of ensembles of orbits.
\par
The key feature entering into the proposed statistical description of
stochastic orbits is the notion of an invariant measure. Mathematically, the
invariant measure corresponds to a probability distribution which, if evolved
into the future using the equations of motion, remains invariant (cf.
Lichtenberg and Lieberman 1992). Applied to galactic dynamics, it corresponds
to a time-independent phase space distribution for stochastic orbits
moving in some specified galactic potential.
\par
There are two important features about the invariant measure which are
relevant in the present context. The first is that, for many time-independent
potentials, generic ensembles of initial conditions corresponding to
stochastic orbits will evolve on relatively short time scales towards an
invariant measure. An initially localised ensemble of stochastic orbits will
disperse and eventually, via a sort of phase mixing, evolve towards a
distribution which fills all of the accessible phase space, with a relative
weight for different phase space regions that is given by the invariant
measure. The second feature is that the invariant measure defines the natural
ensemble of stochastic orbits to consider if one is interested in
time-independent equilibrium configurations.
In particular, it constitutes the natural building block for the construction
of self-consistent models which include stochastic orbits.
\par
If one is interested in models of galaxies that do not exhibit strict
spherical or axisymmetry, it is in general impossible to construct analytic
equilibria. For that reason, one is typically forced to proceed
numerically via Schwarzschild's (1979) method, or some variant thereof.
The idea underlying this method is straightforward, at least in principle.
What one must do is (1) specify some time-independent gravitational
potential $\Phi$, (2) generate a large number of orbits in that potential,
and then (3) select an ensemble of orbits which yields a mass distribution
$\rho$ that generates $\Phi$ self-consistently as a solution to the Poisson
equation, ${\nabla}^{2}\Phi=4{\pi}G{\rho}$. The basic point now is clear.
If the self-consistent configuration is to constitute a true equilibrium,
it must have the property that, when evolved into the future using the
self-consistent equations of motion, its form remains unchanged. However,
when considering stochastic orbits, the only way to insure that this be true
is to demand that the orbits be chosen to sample the time-independent
invariant measure.
\par
As a concrete example, consider a nonintegrable two degree of freedom system
characterised by a time-independent Hamiltonian $H$. Because $H$ is independent
of time, the energy $E$ of any given orbit is conserved, so that evolution is
restricted to a three-dimensional hypersurface of constant $E$. Suppose further
that there exist no additional conserved quantities. Generically, this
constant energy hypersurface will then contain both islands of regular orbits
and a surrounding sea of stochastic orbits. The stochastic regions may be
divided into disjoint regions by invariant tori. Suppose, however, that the
stochastic regions are all connected, so that any given orbit can access all
of the stochastic portions of the constant energy hypersurface. The evolution
towards an invariant measure then corresponds to the fact that a generic
ensemble of initial conditions, corresponding to stochastic orbits of energy
$E$, will evolve towards an invariant distribution ${\G}(E)$, the form of
which depends only on the energy and is independent of all other details.
\par
The two basic questions should now be clear: How rapid and efficient is this
evolution towards the invariant measure, and how should one characterise the
stochastic properties of this invariant measure? Answering the first of these
questions involves probing the rate at which a generic ensemble evolves
towards an ``equilibrium.'' Answering the second involves probing the
properties of that equilibrium.
\par
At least for some model potentials, one knows that generic ensembles of
initial conditions corresponding to stochastic orbits can evidence a
very rapid coarse-grained evolution towards an invariant measure. For
example, Kandrup \& Mahon (1994) have shown that, for the sixth order
truncation of the Toda lattice potential, generic ensembles of fixed energy
$E$ evolve exponentially towards an invariant measure on a time scale which,
in physical units, is substantially shorter than $t_H$. Moreover, they have
shown that the rate at which ensembles of energy $E$ evolve towards the
invariant measure ${\G}(E)$ is directly related to the value of the Liapounov
exponent ${\c}(E)$. There is no guarantee that a similar behavior will be
observed for all potentials. In particular, there may be other systems in
which an invariant measure does not even exist. However, in such cases it
would seem impossible to construct a time-independent self-consistent model
which incorporates stochastic orbits.
\par
Given the importance of the invariant measure ${\G}$, it is natural to search
for useful characterisations of the statistical properties of orbits that
sample ${\G}$. One way in which to characterise this measure is
to probe the overall degree of instability exhibited by orbits within the
ensemble on physically relevant time scales ${\D}t{\;}{\le}{\;}t_{H}$.
Precisely this information is contained within a calculation of short time
Liapounov characteristic numbers
$${\c}({\D}t){\;}{\equiv}{\;}\lim_{{\d}r(0)\to 0}
{\;}{1\over {\D}t}{\;}{\rm log}{\Biggl(}
{{\d}r({\D}t)\over {\d}r(0)}{\Biggr)}. \eqno(2)$$
\par
Given the values of ${\c}({\D}t)$ for all the orbits in the ensemble, one can
of course compute $N({\c}({\D}t))$, the distribution of Liapounov
characteristic numbers, which depends on both the form of the invariant
measure, and hence the energy, and on the duration of the sampling interval.
One would expect that the form of this distribution should depend on the
invariant measure since the avlues of the Liapounov exponents depend on the
energy. Moreover, the distribution must depend on the sampling
interval ${\D}t$ since, as one samples for progressively longer times, one must
eventually converge towards a distribution that is infinitely sharply peaked
about the conventional Liapounov exponent.
\par
Even though this prescription differs substantively from the standard approach
involving asymptotic orbital dynamics, the overall conclusions may not be all
that different. The basic reason for this is that time averages computed for an
individual orbit can coincide with ensemble averages generated from a
collection of orbits, provided that the ensemble is so chosen as to sample
the invariant measure. This would imply in particular that the mean
Liapounov characteristic number ${\overline\chi}$ associated with the
distribution $N({\c}({\D}t))$ should coincide with the standard Liapounov
exponent ${\chi}$, as calculated in the usual way from a single orbit for
very long times. As discussed more carefully in Sections $3$ - $5$, this is
actually true for at least one model potential, namely the truncated Toda
potential (cf. Kandrup \& Mahon 1994).
\par
More generally, one might expect that the information required for the
standard calculation of the Liapounov exponent $\chi$ via a long time
integration actually contains within it the same information as is contained
within the short time distribution $N({\c}({\D}t))$. Given a sequence
of estimates $\{ {\chi}(t_{i})\}$, $(i=1,2,...)$, computed at fixed intervals
$t_{i+1}-t_{i}={\Delta}t$, one can extract a sequence of short time Liapounov
characteristic numbers ${\chi}({\Delta}t_i)$ via the obvious prescription
$${\chi}({\Delta}t_i){\;}{\equiv}{\;}
{{\chi}(t_{i}+{\Delta}t)[t_{i}+{\Delta}t]-{\chi}(t_{i})t_{i}\over {\Delta}t} .
\eqno(3)$$
The critical point then is that, if time and ensemble averages agree,
the distribution of Liapounov characteristic numbers ${\chi}({\Delta}t_i)$
generated in this way should coincide (to within statistical errors) with the
distribution $N({\c}({\D}t))$.
\par
In the remaining sections of this paper, this general picture will be
confirmed in detail for one specific model potential, namely the sixth
order truncation of the Toda lattice. What this means is that the standard
Liapounov exponent $\c$, defined formally as a $t\to\infty$ limit, actually
has physical meaning on short time scales as well. $\c$ measures the mean
stochasticity for an ensemble of orbits that samples the invariant measure.
This implies that earlier work (cf.  Contopoulos \& Barbanis 1989)
tracking the time evolution of Liapounov characteristic numbers for long
times can be exploited to extract information about the short times
distribution $N({\c}({\D}t))$.
\par
It is also clear that the Liapounov exponent $\c$ is only one moment of
$N({\c}({\D}t))$, which can also be supplemented by a consideration of
higher moments. For example, a calculation of the dispersion ${\s}_{\c}$
gives a measure of the degree to which individual Liapounov characteristic
numbers deviate from the mean. Moreover, the overall shape of the distribution
can provide information about ``stickiness'' associated with islands of
regular orbits. More precisely, one knows that stochastic orbits can
sometimes get trapped temporarily in the neighborhood of an island, and
one might expect that, if trapped in that neighborhood, $\c({\D}t)$
will be substantially smaller than the mean value ${\overline\chi}$.
\par
The equivalence of time and ensemble averages discussed above means that
one can hope to apply at least one basic idea from ergodic theory to
systems which contain both regular and stochastic orbits. Ergodic systems
are systems in which the entire phase space is stochastic, and in which
time and ensemble averages agree, provided that the ensemble used to
construct the average is chosen appropriately (cf. Lichtenberg and Lieberman
1992). Thus, e.g., the standard ergodic hypothesis underlying equilibrium
thermodynamics implies that, for an isolated Hamiltonian system, the natural
ensemble corresponds to a microcanonical distribution, i.e., a uniform sampling
of the constant energy hypersurface. Realistic galactic potentials do not seem
to define ergodic systems, since even the most chaotic potentials which have
been envisioned contain at least some regular orbits, and are
characterised by an invariant measure which is not microcanonical (cf. Kandrup
\& Mahon 1994). However, despite deviating from ergodicity in these respects,
they do appear to satisfy the important condition that, for the stochastic
orbits, time and ensemble averages agree, provided that one works with
ensembles that sample the invariant measure.
\par
When considering regular orbits, it is physically well motivated to consider
the properties of individual trajectories since, typically, infinitesimal
changes in initial conditions will not lead to gross qualitative changes in the
subsequent evolution. Thus, in particular, initially nearby trajectories
remain nearby and the overall shape of an orbit is typically stable towards
small perturbations. However, for the case of stochastic orbits this is no
longer true. The fact that these orbits have positive Liapounov exponent
implies that two initially nearby trajectories will typically diverge
exponentially. In this case, it is arguably more appropriate to follow the
evolution of an initial phase space element ${\D}{\bf z}(t_{0})$
than the evolution of a single material point. The obvious fact then is that
the distribution $N({\c}({\D}t))$ provides a characterisation of the average
divergence of nearby trajectories and, as such, information about the
subsequent spreading of an initially localised phase space element.
\par
All of this has focused on the role of Liapounov characteristic numbers in
characterising the behavior of ``equilibrium'' ensembles that sample the
invariant measure. However, one can also consider short time Liapounov
characteristic numbers in the context of the approach towards equilibrium.
Numerically, realisations of the invariant measure can be constructed by
choosing any ``random'' ensemble of initial conditions corresponding
to stochastic orbits and evolving these initial data forward in time until,
statistically, they approach a sampling of a time-independent distribution.
However, while evolving these initial conditions into the future, one can
simultaneously compute short times Liapounov characteristic numbers
${\c}({\D}t)$, which can provide a useful statistical characterisation of
the overall stochasticity of the initial conditions as they evolve towards
the invariant measure.
\vskip .2in
\par\noindent
{\bf 3. EQUIVALENCE OF TEMPORAL AND ENSEMBLE AVERAGES}
\par{\bf AND THE DISTRIBUTION OF SHORT TIME ${\bf{\c}}$'S}
\vskip .1in
\par\noindent
The sixth order truncation of the Toda lattice corresponds to a Hamiltonian
system of the form
$$H={1\over 2}{\Bigl(}p_{x}^{2}+p_{y}^{2}{\Bigr)}+V(x,y), \eqno(4) $$
where
$$V(x,y)={1\over 2}{\Bigl(}x^{2}+y^{2}{\Bigr)}+x^{2}y-{1\over 3}y^{3}
+{1\over 2}x^{4}+x^{2}y^{2}+{1\over 2}y^{4} $$
$$+x^{4}y+{2\over 3}x^{2}y^{3}-{1\over 3}y^{5}
+{1\over 5}x^{6}+x^{4}y^{2}+{1\over 3}x^{2}y^{4}+{11\over 45}y^{6}. \eqno(5)$$
This defines a two degree of freedom system in terms of the canonical pairs
$\{x,p_{x}\}$ and $\{y,p_{y}\}$. The potential $V(x,y)$ can be derived
formally from the true Toda potential
$$V(x,y)={1\over 24}{\biggl\{}
{\rm exp}{\bigl[}2(3)^{1/2}x+2y{\bigr]} +
{\rm exp}{\bigl[}-2(3)^{1/2}x+2y{\bigr]}+{\rm exp}(-4y){\biggr\}}-{1\over 8}
\eqno(6) $$
by a Taylor series in $x$ and $y$, truncated at sixth order.
\par
The Hamiltonian system defined by the true Toda potential (5) is integrable,
but it is believed that all truncations at third and higher order are
nonintegrable. In particular, one knows that the sixth order truncation
is nonintegrable, and, that, for energies above a critical value
$E{\;}{\approx}{\;}0.80$, there exist both regular and stochastic orbits
(cf. Contopoulos and Polymitis 1987).
The Toda lattice and its truncations are all special in the sense that they
manifest a discrete $2\pi/3$ rotational symmetry. One knows (cf.
Udry \& Pfenniger 1988) that special symmetries of this form can have
important effects on the form of the regular orbits and the relative
abundance of regular and stochastic orbits. However, experience
would suggest that, even though breaking this, or any other, symmetry will
typically increase the relative abundance of stochastic orbits, the behavior
of the stochastic orbits in this potential should be qualitatively
similar to stochastic orbits in other potentials with different or less
symmetry.
\par
Having chosen to consider a truncated Toda potential, it were perhaps
appropriate to justify the particular truncation at {\it sixth} order. This
is in fact easily done: If one is concerned with the evolution towards an
invariant measure, one must consider a potential in which the orbits are
confined to a compact phase space region. This is achieved by demanding that
$V(x,y) \to \infty$ as $x$ and $y \to \infty$, which requires an even order
truncation. One also wishes to consider the lowest order truncation possible,
since this will minimise the time required in effecting the numerical
computations. The second order truncation is of course integrable, and hence
unacceptable. The fourth order truncation is not integrable, as proved
by Yoshida et al (1988) and demonstrated explicitly by Udry and
Martinet (1990). However, albeit nonintegrable, this lower order truncation
exhibits relatively little chaos, perhaps because the leading order quartic
terms in the potential (as well as the quadratic terms) are strictly
axisymmetric.
\par
In order to study stochastic orbits in a given potential, one must
first ascertain the location of the stochastic regions as a function of
energy $E$. For sixteen different values of energy between $E=10$ and
$E=200$, surfaces of section were generated, plotting $y$ and $p_y$ at
successive points where randomly chosen orbits pass through the value
$x=0$. This is a useful choice of section because the Toda potential is
symmetric under a reflection $x\to -x$, so that each orbit must repeatedly
intersect the $x=0$ hyperplane. These surfaces of section were then used
to select initial data corresponding to ten different stochastic orbits
for each of the sixteen selected values of the energy.
\par
For each of these ten different choices of initial conditions, Liapounov
exponents were computed in the standard way (cf. Bennetin et al 1976).
This was done by introducing a small initial perturbation ${\d}x=10^{-10}$
into each orbit, continually solving the variational equations for the
perturbation as the unperturbed orbit is evolved, and then renormalising
the evolved perturbation back to a total amplitude
$${\d}z=({\d}x^2+{\d}y^2+{\d}p_x^2+{\d}p_y^2)^{1/2}=10^{-10} \eqno(7) $$
at intervals ${\D}t=10$. Successive computations of ${\c}(t)$ were
made at the same time intervals, with the calculation for each
orbit proceeding for a total time $t=10^4$. The value ${\c}(t=10^4)$
was then interpreted as providing an estimate for the true Liapounov
exponent, which is of course defined only in the $t\to \infty$ limit.
It was found thereby that, for orbits of fixed energy, the value of the
Liapounov exponent is independent of the initial conditions, so that one
can speak of a unique ${\c}(E)$.
\par
The surfaces of section were also used as an aid to construct three localised
ensembles of initial conditions of fixed energy $E$ within the stochastic
phase space regions. Each cell was constructed as follows: One started by
selecting a point $\{y,p_y\}$ on the $x=0$ surface of section which is
displaced significantly from any large islands of regular orbits. This point
was then used to define the center of a cell of specified size ${\D}y$ and
${\D}p_y$. A uniform sampling of this cell via a rectangular grid served to
select $400$ pairs $\{y,p_y\}$, and these pairs were then used to generate an
ensemble of initial conditions $\{x,y,p_{x},p_{y}\}$, setting $x=0$ and
$$p_{x}{\;}{\equiv}{\;}
{\biggl\{}2{\Bigl[}E-V(x=0,y){\Bigr]}-p_{x}^{2}{\biggr\}}^{1/2}>0.
\eqno(8)$$
For energies $E{\;}{\ge}{\;}50$, ${\D}y=0.2$ and ${\D}p_{y}=0.72$. For
lower energies, ${\D}y$ and ${\D}p_{y}$ were chosen to be one half or
one quarter as large.
\par
As described elsewhere (Kandrup \& Mahon 1994), when these ensembles
of initial conditions were evolved into the future, they were found to
evidence a coarse-grained, exponential approach towards an invariant
measure on a time scale $t{\;}{\ll}{\;}100$. For this reason, the collections
of $400$ phase space coordinates at time $t=100$ were taken as constituting
random samplings of the invariant measure ${\G}(E)$. Short time $t=100$
Liapounov characteristic numbers were computed for these random samplings
of the invariant measure, with ${\c}(t)$ for each orbit again being recorded
at ${\D}t=10$ intervals. The resulting distribution of Liapounov
characteristic numbers, $N({\c}(t=100))$, was then analysed to extract the
first and second moments, ${\overline\chi}(t=100)$ and ${\s}_{\c}(t=100)$.
\par
The first striking result derived from such an analysis is that the
mean value ${\overline\chi}$ associated with the short time Liapounov
characteristic numbers coincides, to within statistical uncertainties, with
the long time estimate of the Liapounov exponent ${\c}$. This is
illustrated in Fig. 1, which exhibits the estimated values of the
Liapounov exponents computed in both ways. To generate this Figure,
the ten long time estimates at each energy $E$ were first averaged
together to obtain a mean Liapounov exponent. The values obtained thereby
were then plotted as small triangles, with error bars that represent the
standard deviations associated with the means. The three larger diamonds for
each value of $E$ represent calculations of the mean values
${\overline\chi}$ associated with the short times distributions,
$N({\c}(t=100))$, generated from ensembles of $400$ orbits.
\par
The second significant result derived from this analysis is that the long
time calculation of ${\c}$ for a single stochastic orbit actually contains
within it the same information as the distribution of short time
Liapounov characteristic numbers derived for an ensemble of orbits.
To extract this information, the values of ${\chi}(t)$ for a single orbit
were partitioned as in Eq. (3) into intervals ${\D}t=100$ to extract
a different set of short time estimates ${\c}({\D}t_i)$. The key point then
is that the distribution of ${\c}({\D}t_i)$'s generated in this way is almost
identical to $N({\c}({\D}t)$, the distribution of short time ${\c}$'s
generated from the ensembles of orbits that sample the invariant measure.
\par
Fig. 2 illustrates the forms of these distributions for four different
values of energy, namely $E=150$, $75$, $30$, and $20$. These distributions
were generated by binning the data into intervals ${\delta}{\c}=0.05$ and
normalising the resulting binned distribution so that the most populous
bin is assigned the value $N({\c})=1$.
\par
The first obvious point to be inferred from this Figure is that, as noted
already, except for some small differences at very low values of ${\c}$
the two distributions are essentially the same. The second point is that, at
least at relatively high energies, both distributions are extremely well
approximated by a Gaussian form. However, for energies below a value
$E{\;}{\approx}{\;}75-100$, the distribution begins to acquire a statistically
significant short ${\c}$ tail. When the energy is decreased further to a
value as small as $E{\;}{\approx}{\;}40-60$, this tail then acquires a
secondary peak. However, at very low energies, $E{\;}{\approx}{\;}10-20$,
the two peaks eventually merge into a single distribution which deviates
significantly from a Gaussian.
\par
The observed structure at small values of ${\chi}$
seems to be associated with the increasing predominance of regular orbits
at lower energies. It is well known (cf. MacKay et al 1984a,b, Lau et al 1991)
that stochastic orbits that stray too close to islands of regularity
tend to get trapped for relatively long periods of time around
these islands, and it would appear that, while in these regions,
their Liapounov characteristic numbers are substantially reduced.
\par
The differences between the two distribution at small values of
${\chi}<0.1-0.15$ arise because the alleged sampling of the
invariant measure is contaminated by a small number of regular orbits. In
constructing these samplings, one selected an initial phase space
region which seemed, at least superficially, to be comprised completely
of stochastic orbits. However, these stochastic regions actually
contain a nonzero measure of regular orbits embedded in the stochastic
sea.
\par
The basic conclusion, therefore, is that,
because of the equivalence of time and ensemble averages, an evaluation
of ${\c}(t)$ for a single orbit over long times contains the same information
as the distribution of short time Liapounov characteristic numbers associated
with an ensemble of orbits that samples the invariant measure.
\vskip .2in
\par\noindent
{\bf 4. LIAPOUNOV CHARACTERISTIC NUMBERS AND THE}
\par
{\bf  APPROACH TOWARDS AN INVARIANT MEASURE}
\vskip .1in
\par\noindent
The preceding section focused on the form of the distribution of
Liapounov characteristic numbers associated with an initial ensemble
that samples the invariant measure. However, as discussed in Section 2,
one can also consider the distribution $N({\c})$ associated with other
ensembles as well. In particular, it is natural to compute the distribution
of Liapounov characteristic numbers for the initially localised ensembles
considered in Section 3 as they evolve towards the invariant measure.
Doing this enables one to understand how the evolution towards an
invariant measure correlates with the overall stochasticity of the
orbits, a problem already considered in a slightly different context by
Kandrup \& Mahon (1994).
\par
The results for one particular energy, namely $E=50$, are illustrated
in Fig. 3. The data plotted in this Figure were generated by selecting
the three initially localised ensembles discussed in Section 3,
and computing ${\c}(t)$ for each of the $400$ orbits in each ensemble
at $t=10$ intervals for a total time of $t=200$. The resulting ${\c}$'s
were then partitioned into collections of short time estimates
${\c}({\D}t_i)$ at ${\D}t=10$ intervals. Fig. 3 exhibits the mean
${\overline\chi}({\D}t_i)$ and the associated dispersions
${\s}_{\c}({\D}t_i)$ for these three different ensembles.
\par
In this Figure, the first interval ${\D}t=100$ is interpreted as corresponding
to the period during which the ensemble evolves towards the invariant measure.
The phase space coordinates at $t=100$ correspond to the sampling of the
invariant measure used for the computations in Section 3, and the remaining
${\D}t=100$ are thus interpreted as corresponding to the evolution of an
initial ensemble that samples the invariant measure.
\par
 From this Figure several conclusions are apparent. The first is that
even at very early times, when the ensemble deviates significantly from a
sampling of the invariant measure, the values of ${\overline\chi}({\D}t_i)$
and ${\s}_{\c}({\D}t_i)$ do not deviate all that much from the
``equilibrium'' values associated with the invariant measure. At early times,
${\overline\chi}$ tends to be somewhat larger than the value
associated with the invariant measure. This is easily understood by observing
that the initial ensemble was located in a phase space region particularly far
from any large islands of regular orbits where, overall, one might expect a
higher degree of stochasticity. At early times, one also observes that
${\s}_{\c}$ is somewhat smaller than the value associated with the
invariant measure. This is again easily understood by observing that
the initial ensemble is constructed from a collection of phase space
points that are relatively close together and significantly displaced
from any large islands.
\par
It is also apparent that, by a time $t=100$, ${\overline\chi}$ and
${\s}_{\c}$ have asymptoted towards constant values, this corroborating
the expectation that the phase space coordinates at $t=100$ can be
interpreted at least appropriately as constituting a random realisation
of the invariant measure. For higher values of energy, the approach
towards the invariant measure is even more rapid. Indeed, as discussed
more extensively in Kandrup \& Mahon (1994), there exists a direct
one-to-one correlation between the exponential approach towards an
invariant measure and the value of the Liapounov exponent. Specifically,
as was demonstrated in especial detail for energies
$10{\;}{\le}{\;}E{\;}{\le}{\;}75$, increasing $E$ increases both (1)
the value of the Liapounov exponent  ${\c}(E)$ associated with stochastic
orbits of that energy {\it and} (2) the exponential rate ${\L}(E)$
associated with the approach towards the invariant measure, in such a
fashion that the ratio ${\cal R}(E){\;}{\equiv}{\;}{\L}/{\c}$ is
approximately constant, independent of energy. The fact that the mean
${\overline\chi}(t)$ associated with the evolution of the initially localised
ensemble is rather close in value to the Liapounov exponent ${\c}$, as defined
in a $t\to\infty$ limit, explains why there can exist a direct connection
between the ``equilibrium'' ${\c}$ and the ``nonequilibrium'' evolution
towards an invariant measure.
\par
Finally, it should perhaps be noted that the first points in Fig. 3
at $t=10$ may be somewhat suspect. Specifically, the values of ${\c}(t=10)$
may be significantly influenced by the particular choice of initial
perturbation, namely ${\d}x=10^{-10}$ and ${\d}y={\d}p_x={\d}p_y=0$.
\vskip .2in
\par\noindent
{\bf 5. THE EFFECTS OF A VARYING SAMPLING TIME}
\vskip .1in
\par \noindent
In this Section, attention focuses on how the form of the distribution of
short time ${\c}$'s extracted from a single long time integration, or
generated from an ensemble that samples the invariant measure, varies as a
function of the sampling interval ${\D}t$. One anticipates physically that,
as the length of the sampling interval increases, the mean Liapounov
characteristic number ${\overline\chi}$ will provide an increasingly better
characterisation of the overall stochasticity, and that the dispersion
${\s}_{\c}$ associated with the distribution will tend to zero for
${\D}t\to\infty$.
\par
Given a long time calculation of ${\c}(t)$ for a single orbit, one can
extract short time ${\c}({\D}t_i)$'s for various choices of sampling
interval and compare the results. Fig. 4a exhibits the values of
${\c}({\D}t_i)$ for one long time integration of an orbit with $E=150$,
for a sampling interval ${\D}t=10$. It is clear that ${\c}({\D}t_i)$
shows rapid large amplitude variability, so that the degree of
stochasticity exhibited by the orbit at time $t$ may be substantially
different at times $t{\;}{\pm}{\;}{\D}t$. However, if one considers
a longer sampling interval, the variability is substantially reduced.
This is illustrated in Fig. 4b, which analyses the same data for
sampling intervals ${\D}t=100$. This Figure was generated from Fig.
4a by taking time averages of successive collections of ten intervals
with ${\D}t=10$.
\par
The differences between Figs. 4a and 4b can be quantified by specifying
the maximum and minimum values of ${\c}({\D}t_{i})$, as well as
the standard deviation about the mean. For the shorter time sampling,
the minimum and maximum values of ${\c}$ are $-0.125$ and $1.670$, and
the standard deviation is $0.2997$. For the longer time sampling, the
minimum and maximum values are $0.078$ and $1.053$, and the standard
deviation is approximately half as large, namely $0.1449$.
However, even though the longer time sampling is smoother, nontrivial
structures are still observed. In particular, there is a significant
dip in the value of ${\c}$ between $t=4000$ and $4300$ which is
observed in both figures, where ${\c}({\D}t_{i})$ decreases appreciably
to a value $<0.1$.
\par
To further elucidate these sorts of differences, one can also examine
the form of the distribution $N({\c}({\D}t))$ as a function of the
sampling interval. Figures 5a and 5b illustrate the form of $N({\c}({\D}t))$
for two different energies, $E=150$ and $50$, and three different
sampling intervals, ${\D}t=10$, $40$, and $100$. These figures were
constructed by partitioning each of the ten $t=10^4$ calculations
into a collection of intervals of fixed length ${\D}t$, combining
the data for all ten orbits, and constructing a binned distribution
with ${\D}{\c}=0.05$. For both values of the
energy, it is clear that the width of the distribution decreases as
the sampling interval becomes longer. It is also clear that, for each
choice of sampling interval, the relative width of the low energy
distribution is slightly larger than that of the high energy
distribution. For $E=150$, the distribution is essentially Gaussian
for the two larger sampling intervals, whereas, for $E=50$ one sees
indications of a low ${\c}$ tail.
\par
It is natural to ask whether the systematic decrease in the width of
the distribution, as probed by the dispersion, is exponential in
time, or whether it is better fit by a power law. The answer is
that, for all values of the energy, the dispersion is well fit by
a power law ${\s}_{\c}{\;}{\propto}{\;}{\D}t^{-p}$, where $p$ is a
positive constant. The goodness of fit is illustrated in Fig. 6 for
three different energies, $E=50$, $125$, and $175$.
\par
One might also ask how the value $p$ depends on the energy $E$.
It is difficult to extract a precise estimate of $p$ for any given
energy because the best fit value depends sensitively on whether or
not there exists one or two initial conditions where the orbit spends
a large amount of time in a region of especially low ${\c}$.
However, one can still extract a best fit value of $p$ for each
energy. The result of such an analysis is exhibited in Fig. 7.
Because of the large scatter in this plot, it is difficult to extract
a precise functional form for $p(E)$. However, two trends are
unambiguous: First, it is clear that, overvall, the value of $p$ increases
with increasing energy, and second, it would appear that the value
of $p$ approaches $0.5$ at large energy.
\par
A value of $p=0.5$ is relatively easy to explain. The distribution
$N({\c}({\D}t))$ for large ${\D}t$ can be viewed as a convolution
of a large number of distributions $N({\c}({\D}t_i))$ for shorter
time intervals ${\D}t=10$. Suppose however, that the values of
${\c}({\D}t_{i})$ for successive ${\D}t=10$ intervals are completely
uncorrelated. In this case, the central limits theorem guarantees that
the long time distribution will converge towards a Gaussian with a
dispersion that scales as $k^{-1/2}$, where $k$ denotes the total number
of short time intervals. The fact that this number grows linearly in
time would then imply a dispersion that scales as $t^{-1/2}$. If these
successive intervals are not completely uncorrelated, then the dispersion
should decrease more slowly. Fig. 7 therefore indicates that, at low
energies, there is substantial correlation between successsive ${\D}t=10$
intervals, but that this correlation decreases for higher energies.
\vskip .2in
\par\noindent
{\it Acknowledgments.} The authors acknowledge useful discussions with Robert
Abernathy, Salman Habib, Edward Ott, and Daniel Pfenniger. Some of the
computations were
facilitated by computer time made available through the Research Computing
Initiative at the Northeast Regional Data Center (Florida) by IBM Corp. HEK
was supported in part by the NSF grant PHY92-03333. MEM was supported by the
University of Florida as a Postdoctoral Research Associate.
\vfill
\eject
\par\noindent
{\large\bf References}
\vskip .2in
\par \noindent
Bennetin, G., Galgani, L., Strelcyn, J.-M., 1976, Phys. Rev. A 14, 2338
\par\noindent
Chirikov, B., 1979, Phys. Repts. 52, 265
\par \noindent
Contopoulos, G., Barbanis, B., 1989, A\&A 222, 329
\par \noindent
Contopoulos, G., Grosbol, P., 1989, A\&A Rev. 1, 261
\par \noindent
Contopoulos, G., Polymilis, C., 1987, Physica D 24, 328
\par \noindent
H\'enon, M., Heiles, C., 1964, A\&A 69, 73
\par \noindent
Kandrup, H. E., Mahon, M. E., 1994, Phys. Rev. E, submitted
\par\noindent
Lau, Y.-T., Finn, J. M., Ott, E. 1991, Phys. Rev. Lett. 66, 978
\par\noindent
Lichtenberg, A. J., Lieberman, M. A., 1992, Regular and Chaotic Dynamics,
Springer,
\par Berlin
\par\noindent
MacKay, R. S. Meiss, J. D. Percival, I. C. 1984a, Phys. Rev. Lett. 52, 697
\par\noindent
MacKay, R. S. Meiss, J. D. Percival, I. C. 1984b, Physica D 13, 55
\par \noindent
Martinet, L., Pfenniger, D., 1987, A\&A 173, 81
\par\noindent
Pesin, Ya. B., 1977, Russ. Math. Surveys 32, 55
\par \noindent
Schwarzschild, M., 1979, ApJ 232, 236
\par \noindent
Toda, M., 1967a, J. Phys. Soc. Japan 22, 431
\par \noindent
Toda, M., 1967b, J. Phys. Soc. Japan 23, 501
\par\noindent
Udry, S., Martinet, L, 1990, Physica D 44, 61
\par \noindent
Udry, S., Pfenniger, D., 1988, A\&A 198, 135
\par\noindent
Yoshida, H., Ramani, A., Grammaticos, B., 1988, Physica D 30, 151
\vfill
\eject
\par\noindent
{\bf Fig. 1}. Liapounov exponents as a function of energy $E$ estimated in two
different ways. The small triangles represent ${\chi}$, calculated in the
usual way by following ten different orbits for a total time $t=10^{4}$. The
error bars represent the associated standard deviations. The larger diamonds
represent calculations of the mean ${\overline \chi}(t)$ for
ensembles of $400$ orbits over a period $t=100$. 
\vskip .1in
\par\noindent
{\bf Fig. 2}. (a) The binned distribution of Liapounov characteristic numbers
${\chi}({\D}t)$ for ${\D}t=100$, calculated in two different ways for orbits
with $E=100$. The solid
curve represents the distribution obtained from an initial ensemble that
sampled the invariant measure. The dashed curve represents the distribution
obtained from partitioning the curve ${\chi}(t)$ for $t=10000$ into $100$
segments of length ${\D}t=100$, and construction ${\chi}({\D}t_{i})$ for each
segment. (b) The same for $E=75$. (c) The same for $E=30$. (d) The same for
$E=20$. 
\vskip .1in
\par\noindent
{\bf Fig. 3}. The mean ${\overline \chi}({\D}t_{i})$, and the associated
dispersion ${\s}_{\c}({\D}t_{i})$, computed at ${\D}t=10$ intervals for a
total time $t=200$ for three initially localised ensembles of $400$ stochastic
orbits with $E=50$. The evolved phase space coordinates at $t=100$ were
interpreted as constituting random realisations of the invariant measure, and
used as initial data for the calculations described in Section 3. The solid
lines represent the values of ${\overline\chi}$ and ${\s}_{\c}$ associated
with the invariant measure. 
\vskip .1in
\par\noindent
{\bf Fig. 4}. The Liapounov characteristic number ${\chi}(t)$ for one orbit
with $E=100$ partitioned into a collection of short times estimates
${\chi}({\D}t)$, with (a) ${\D}t=10$ and (b) ${\D}t=100$. 
\vskip .1in
\par\noindent
{\bf Fig. 5}. (a) The distribution of short times ${\chi}({\D}t)$ for $E=150$,
extracted from the long time ${\chi}(t)$. The solid curve represents a
sampling interval ${\D}t=100$, the dot-dashed line ${\D}t=40$, and the dashed
line ${\D}t=10$. (b) The same for $E=50$. 
\vskip .1in
\par\noindent
{\bf Fig. 6}. The dispersion ${\s}_{\c}$ as a function of sampling interval
${\D}t$ for energies $E=50$ (bottom curve), $125$ (middle curve), and $175$
(top curve). The solid curves represent least squares power law fits
${\s}_{\c}({\D}t){\;}{\propto}{\;}{\D}t^{-p}$. The curves for $E=50$ and
$E=175$ were displaced respectively upwards and downwards by ${\s}_{\c}=0.125$
to provide a less cluttered diagram. 
\vskip .1in
\par\noindent
{\bf Fig. 7}. The exponent $p$ associated with the least squares fit
${\s}_{\c}{\;}{\propto}{\;}{\D}t^{-p}$, plotted as a function of energy $E$.
\vfill
\eject
\end{document}